\documentclass[journal]{IEEEtran}

\usepackage{graphicx}
\usepackage[english]{babel}
\usepackage{epsfig,amssymb,amsmath}
\usepackage{verbatim}
\usepackage{subfigure}
\usepackage{bm}

\begin{document}

\title{Engineering interband tunneling in nanowires with diamond cubic or zincblende crystalline structure based on atomistic modeling}

\author{
\IEEEauthorblockN{Pino~D'Amico\IEEEauthorrefmark{1}\IEEEauthorrefmark{2},
Paolo~Marconcini\IEEEauthorrefmark{1},
Gianluca~Fiori\IEEEauthorrefmark{1}, 
Giuseppe Iannaccone\IEEEauthorrefmark{1}}\\
\IEEEauthorblockA{\IEEEauthorrefmark{1} Dipartimento di Ingegneria dell'Informazione,
Universit\`a di Pisa, Via Caruso, 56126 Pisa, Italy.}\\
\IEEEauthorblockA{\IEEEauthorrefmark{2}Current Affiliation: S3 CNR-NANO, Via Campi 213/A, I-41125 Modena, Italy.\\
email: pino.damico@nano.cnr.it}
}


\maketitle

\begin{abstract}

We present an investigation in the device parameter space of band-to-band tunneling in nanowires with a diamond cubic or zincblende crystalline structure. 
Results are obtained from quantum transport simulations 
based on Non-Equilibrium Green's functions with a tight-binding atomistic Hamiltonian.
Interband tunneling is extremely sensitive to the longitudinal electric field, to the nanowire cross section, through the gap, and to the material. We have derived an approximate analytical expression for the transmission probability based on WKB theory and on a proper choice of the effective interband tunneling mass, which shows good agreement with results from atomistic quantum simulation.

\end{abstract}


\section{Introduction}

Conventional field-effect transistors (FETs) are based on the modulation of  thermionic injection of charge carriers from the source by tuning the potential barrier between channel and source via the voltage applied on the gate electrode.
A crucial parameter in FETs is the subtreshold swing $S$ - defined as the gate voltage variation required for a tenfold increase of the drain current.  In the ideal case of perfect electrostatic control, when the potential in the channel exactly follows the gate voltage,
$S$ is limited by the tail of the Maxwell-Boltzmann distribution to $k_B T \ln 10$, where $k_B$ is Boltzmann's constant and $T$ is the 
temperature. 

At room temperature, the intrinsic limitation on $S$ is therefore $S \ge 60$~mV/decade, which represents a real threat to
continuous integration of semiconductor technology, due to conflicting constrains. Indeed, supply voltage ($V_{DD}$) reduction must accompany device size scaling to keep power consumption per unit area ($\propto V_{DD}^2$) under control, and digital logic still requires the ratio of the current of the device in the on state to be at least $10^4$ larger than than in the off state, effectively requiring the threshold voltage $V_{th} \ge 4S $.

A promising option to reformulate this trade-off is to use a device based on a modified operating principle such as the tunnel FET (TFET)
\cite{Hansch00tsf}. In a TFET the gate voltage allows to modulate the band-to-band tunneling (BTBT) current between source and channel,
by modulating the width of the energy window in which tunneling can take place. In this way, both ends of the allowed energy window for transport are defined by conduction and valence band edges, and the previously discussed 
limitation on $S$ is therefore removed, allowing further reduction of $V_{DD}$. Several experimental and theoretical works have been dedicated to explore various options for tunnel FETs 
\cite{Nirschl04iedm,Bhuwalka05ieee,Choi07edl,Boucart07ieee,Fiori09edl}.

Two aspects are important for good performance of TFETs. On the one hand, currents in the on state must be large
for obtaining small delay times, shifting one's preference to lower gap semiconductors, when BTBT can be large. On the other hand,
electrostatic control of the potential in the channel via the gate voltage must be ideal, not to unduly lose the advantage on $S$, 
as can be obtained in ultra-thin-body or gate-all-around devices. 

In this work, we investigate the dependence of band-to-band tunneling 
in the parameter space, considering different semiconductor materials, 
nanowire diameters, and electric fields in the interband tunneling barrier. 
Our focus is understanding their role in determining 
the achievable BTBT current and to allow comparison of different material 
systems and channel geometry.
The information extracted at the atomistic level provides relevant and reliable
quantities to be used, within a multi-scale approach, at a higher level 
of abstraction, allowing a faster, but still accurate analysis.

As in~\cite{Luisier06prb}, the approach is based on an atomistic $sp^3d^5s^*$ 
Hamiltonian for diamond cubic or zincblende crystals. 
The spin-orbit interaction is not included in our model because it has been shown that it provides only minor contribution to the final current~\cite{Luisier10jap}. 
We have also verified its effect on the band profiles to be sure that the effect is not crucial.
In particular, we focus on germanium and indium arsenide - which are interesting candidates for TFETs for their relatively small gap as bulk materials - and on silicon, as a reference material for maximum ease of integration with CMOS technology. 
Other materials as GaAs can be investigated within the same computational scheme.
Nanowire transport is computed in the framework of Non-Equilibrium Green's functions (NEGFs).  
Finally, we propose a simple analytical expression derived from the Wentzel-Kramers-Brillouin (WKB) approximation with an ad-hoc tunneling effective mass, that is able to quantitatively reproduce results from atomistic quantum transport simulations and can be used for device modeling at a circuit level.

\section{Results and Discussion}

\begin{figure}[htbp]
\centering
\includegraphics[width=80mm]{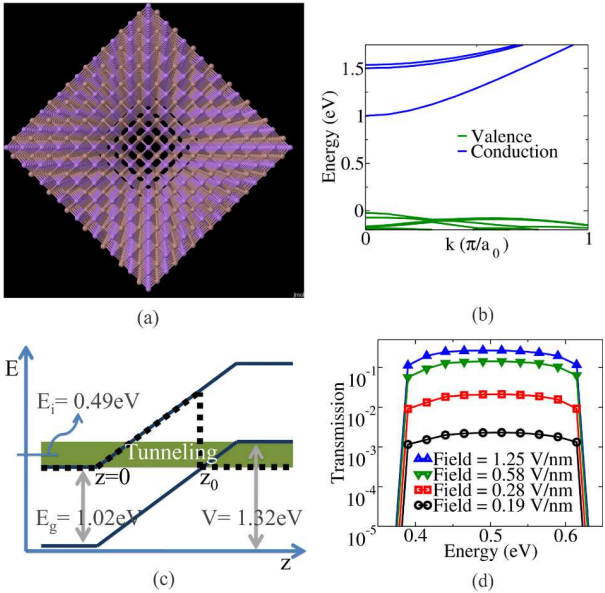}
\caption{Example of the investigated structures and its main  properties: InAs squared nanowire with diameter of 3.4~nm:
(a) lattice structure of the (001) plane transversal to the tunneling direction of the wire; (b)  band structure of the nanowire;
(c) band edge profile considered for the calculation of the tunneling coefficient; the tunneling energy window is highlighted;
(d) transmission coefficient in the tunneling window as a function of energy for different values of the longitudinal electric field.}
\label{graph1}
\end{figure}


In Fig.~\ref{graph1}(a) we show an example of a transversal cross section of a zincblende nanowire with the axis 
oriented in the [001] direction.
We use a tight-binding Hamiltonian with a $sp^3d^5s^*$ first-nearest neighbour representation, where 10 atomic orbitals 
are considered (spin orbit coupling is neglected). 
This representation is extensively employed in modeling semiconducting nanowires and it is the most used for refined numerical calculations \cite{Slater54pr, Boikin02prb, Boikin04prb}.
It also gives quantitative agreement with experimental results \cite{Ganapathi10apl}.


We have developed a tool  to handle generic Hamiltonians with an arbitrary number
of atomic orbitals. In order to speed-up the calculation of the transmission probabilities, 
we have optimized the tool exploiting the four-atomic-layer periodicity
of zincblende materials in the [001] direction.


The computation of  transmission  is based on Non-Equilibrium Green's functions, 
using a recently developed closed-form method for the calculation of the lead self-energies~\cite{Wimmer09phd}.
Due to the complexity of the investigated structures, a closed-form self-energy scheme was required 
to reduce the computing time with respect to iterative procedures. In particular, the adopted procedure has been demonstrated to be ten times faster than the Sancho-Rubio iterative method \cite{Sancho85jpf}.

The transmission probability is obtained with the well known formula
\begin{equation}
T = Tr[\mathbf{G^r} \cdot \mathbf{\Gamma_L} \cdot \mathbf{G^a} \cdot \mathbf{\Gamma_R}],
\end{equation}
where $\mathbf{G^{r,a}}$ are the retarded and advanced Green's function matrices 
and $\mathbf{\Gamma_{L,R}}$ represent the tunneling-rate matrices for the left and right lead, respectively.
A detailed description of the NEGF transport theory can be found in the literature~\cite{Datta,Ryndyk09sscp}

To investigate the possible use of nanowires with zincblende crystalline structure as channels for tunnel FETs
we have computed the transmission coefficient due to band-to-band tunneling.
We first compute the dispersion relations of the nanowire under investigation as shown in Fig.~\ref{graph1}(b), in order to extract the energy gap.
Then, even if the potential in an interband tunneling barrier is not fully linear, we assume a constant electric field as in Fig.~\ref{graph1}(c), to facilitate comparison among different materials and geometry, and to enable the description of interband tunneling in terms of local physical quantities, useful for modeling tools at a higher level of abstraction.
We apply a potential energy drop larger than $E_{g}$ in order to open an inter-band tunneling window of 0.3~eV.
This ensures limited mismatch and a transmission probability almost independent of energy, as shown in Fig.~\ref{graph1}(d), 
where trasmission probability exhibits a plateau in the interband tunneling energy window.
In the following analysis, we use the value of the transmission coefficient in such plateau 
as a metrics for comparing interband tunneling among different materials and as a function of the applied
longitudinal field.


\begin{figure}[t]
\centering
\includegraphics[width=80mm]{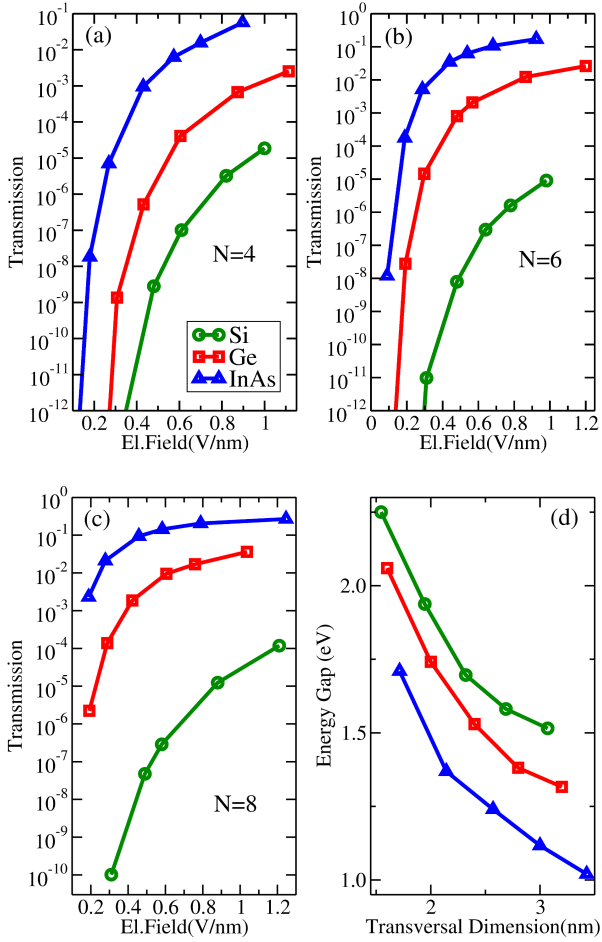}
\caption{ The plateau of transmission probability as a function of the longitudinal electric field is shown for nanowires
along the [001] direction, 
with square cross section of side $l=a_0 \frac{\sqrt{2}}{2}N$ with $N=4$ (a), $N=6$ (b) , $N=8$ (c), where $a_0$ is the lattice constant ($a_0 = 5.43, \,5.65, \,6.05$\AA) for Si, Ge, and InAs, respectively).
In every plot the transmission coefficient is shown for three different materials: Si, Ge and InAs.
(d) dependence of the energy gap on $l$, for the different materials, extracted from the dispersion relation.}
\label{graph2}
\end{figure}

Nanowire tunnel FETs pose different tradeoffs: a smaller nanowire cross section implies a better electrostatic control, and therefore a better subthreshold behavior. On the other hand, it also implies a larger gap, and therefore a smaller tunneling current in the on state.
To explore this tradeoff from a quantitative point of view for different  materials, we consider nanowires in which propagation occurs along the [001]  direction, and the lateral surfaces are on [110] planes, as shown in Fig.~\ref{graph1}(a).

The side of the square cross section $l$ is determined by an integer parameter N according to the formula $l=a_0 \frac{\sqrt{2}}{2}N$  
where $a_0$ is the lattice constant for the material under investigation
($a_0 = 5.43, \,5.65, \,6.05$\AA\ for Si, Ge and InAs, respectively).
In Fig.~\ref{graph2} we show results for $N=4,6,8$, corresponding to a side $l$ between roughly 1.5~nm and
3~nm depending on the material.
As can be seen in Fig.~\ref{graph2}(d), with increasing cross section area the gap rapidly decreases, causing a very steep increase of the tunneling probability. In addition, as can be seen from Figs.~\ref{graph2}(a)-(c) the transmission coefficient increases by 
several orders of magnitude for an increase of the longitudinal electric field of just one order of magnitude.
In the case of InAs, in particular, for the nanowire with $l \approx 3$~nm, the transmission probability can be very high, opening the possibility for very large ON currents.

\begin{figure}[t]
\centering
\includegraphics[width=80mm]{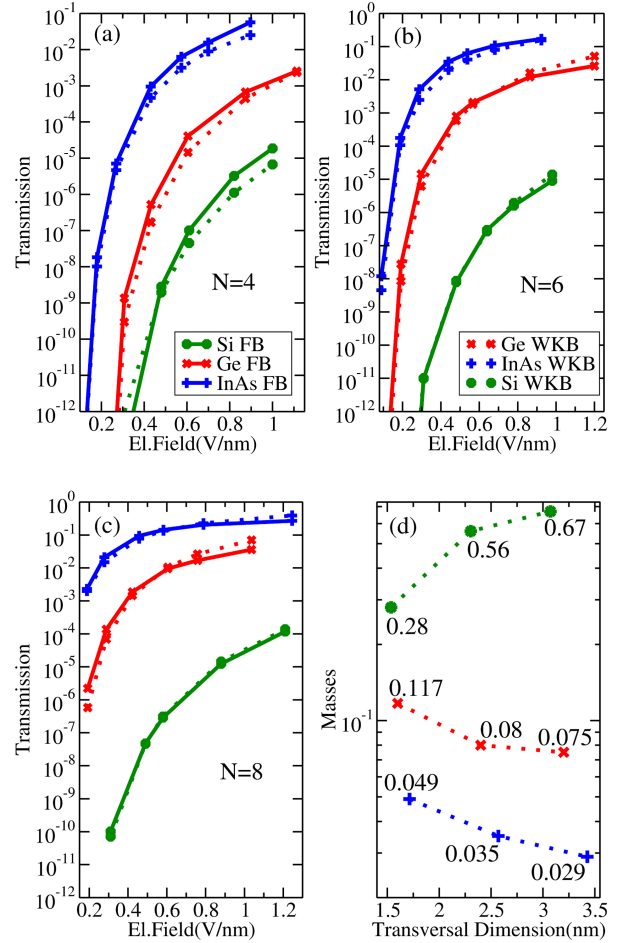}
\caption{Comparison between the transmission probability plateau obtained from the atomistic NEGF calculation (solid lines) and the WKB results (dotted lines).
Nanowires with three different cross section sides $l$ are considered: $l=a_0 \frac{\sqrt{2}}{2}N$ with $N=4$ (a), $N=6$ (b), $N=8$ (c). The values of interband tunnelling effective mass used for the WKB approximation are shown in (d);
they have been extracted from the energy-band profile of the corresponding atomistic calculation for InAs and as a fitting parameter for Ge and Si.}
\label{graph3}
\end{figure}

As we pointed out before, full band calculations, as the ones presented here,
are computationally very demanding for diamond cubic and zincblende nanowires.
It is therefore interesting to verify whether simplified expressions are reliable in predicting the behavior of the tunneling coefficient.
In Fig.~\ref{graph3} a comparison between the transmission probability plateau obtained from the atomistic NEGF calculation and the WKB results is shown. The WKB transmission coefficient is obtained with the formula
\begin{equation} \label{WKB}
T_{\rm WKB} = \exp\left[-2\int_{0}^{z_0}{\sqrt{\frac{2m}{\hbar^2}eFz}\,dz}\right],
\end{equation}
where $F$ is the longitudinal electric field and $m=m_0\cdot m_*$ is the ``interband tunnelling'' effective mass in the 
plotted in Fig.~\ref{graph3}(d) ($m_0$ is the electron mass at rest). 

The potential profile is linear with the slope given by the applied electric field, as shown in Fig.~\ref{graph1}(c).
The integration region is determined by $z_0 = E_{g}/F$.
After few manipulations the formula (\ref{WKB}) can be written as follows:
\begin{eqnarray} \label{WKB_simple}
T_{\rm WKB} \!\!\!\!&=&\!\!\!\!
\exp\left[-\frac{4\,\sqrt{2}}{3}\,\frac{m_0^{\frac{1}{2}}\,e^{\frac{1}{2}}}{\hbar \,\, 10^9}\,
\frac{{E_g}^{\frac{3}{2}}\,{m_{*}}^{\frac{1}{2}}}{F}\right]\nonumber\\
\!\!\!\!&\approx&\!\!\!\!
\exp\left[-\frac{20}{3} m_{*}^{\frac{1}{2}}\frac{E_{g}^{\frac{3}{2}}}{F}\right],
\end{eqnarray}
where $E_g$ is expressed in eV and $F$ in V/nm (this equation coincides with the Eq.~(6) of Ref.~\cite{Ionescu} if $E_g$ is expressed in Joule and $F$ in V/m).

The interband tunnelling effective mass is ``effective'' in the sense that - used in approximating the energy dispersion relation in the gap (i.e. for imaginary wave vectors) - enables to obtain results in agreement with full band quantum simulations. According to definition, it is therefore loosely related to the effective mass in the common definition, which is used for approximating the energy dispersion relation near the band edges, i.e. to describe transport near at the band edge.

Indeed, with (\ref{WKB_simple}) we fit the numerical results for InAs and Ge, but we have to choose  the effective masses in two different ways:
for InAs, which exhibits two clear well-separated conduction and valence bands, the ``interband tunnelling'' effective mass is well approximated by the effective mass in the first conduction band. On the other hand, for Ge and Si the interband tunnelling effective mass must be obtained as a fitting parameter, because multiple quasi-degenerate conduction and valence bands are present. The values of the mass are shown in Fig.~\ref{graph3}(d).
For Si, one synthetic effective mass is not sufficient to reproduce the atomistic NEGF results for the tunneling probability: we need to modify the functional dependence of the WKB transmission probability on the electric field using the formula
\begin{equation} \label{WKB_silicon}
T_{\rm WKB}^{\rm Si} = \exp\left[-\frac{20}{3} m_{*}^{\frac{1}{2}}\frac{E_{g}^{\frac{3}{2}}}{F^{0.71}}\right].
\end{equation}
and to extract the interband tunnelling effective mass as a fitting parameter (values shown in Fig.~\ref{graph3}(d).
As can be seen, the simplified formulas of Eqs.~(\ref{WKB_simple},\ref{WKB_silicon})
reproduce quite accurately full-band calculations for InAs, Ge and Si respectively. 
We want to stress two issues: i) the agreement between WKB and full-band calculations is obtained for a range of wire diameters reaching 3.4 nm for InAs, that is in line with state of the art calculations for this structures and for $sp^3d^5s^*$ models \cite{Luisier10jap}; ii) the agreement is reached for a range of electric field of up to 1 V/nm which is already very high for nanowires.
A more complex implementation of the WKB approximation could in principle be used to obtained better accuracy \cite{Luisier10jap}, 
but our results show that the simplest version can still provide a very good quantitative agreement.
In addition, the provided simple expression can be easily included at a higher
level of abstraction in circuit simulators, or in Monte Carlo codes in order
to provide the generation rate in correspondence of the tunneling barrier.
Therefore, first-principle numerically accurate calculations can be reproduced by simple analytical formulas, with
effective masses extracted from the band profile of the atomistic structure or from fitting procedures.
This makes self-consistent calculations of currents a much less demanding computational task compared to full-band NEGF calculations.

We also want to underline that detailed simulations~\cite{Luisier10jap,Koswatta} show that electron-phonon coupling
only affects current by at most a factor of two, which is a relatively small factor considering that, in the range of electric field considered, the current varies by several orders of magnitude. Indeed, BTBT current is so sensitive to the electric field that - for a given current - coherent transport would overestimate the electric field in the barrier by only few percents. Since this is well below the accuracy with which the electric field in the interband tunneling barrier of an actual device is known, the inclusion of electron-phonon coupling would not change our conclusion.
\section{Conclusion}

We have investigated diamond cubic or zincblende nanowires for different materials and for different 
cross sections up to a side of $3.4$~nm using atomistic Hamiltonians based on $sp^3d^5s^*$ tight-binding representation.

The extreme sensitivity of the tunneling probability to the electric field and to the cross section side provides a hint of the 
sensitivity of the tunnel FET ON current to the variability of the nanowire cross section and to the presence of random charged defects in the nanowire region where interband tunneling takes place.
We have shown that tunneling probabilities between 0.01 and 0.1 - suitable for achieving reasonable ON currents in TFETs - can be obtained with longitudinal electric field close to 1~V/nm (still smaller than the dielectric breakdown field) in InAs for $l > 1.5$~nm and for Ge for $l>3$~nm. We have also shown that full-band calculations are quantitatively reproduced by simple formulas derived from the WKB approximation, considering an ad hoc interband tunneling mass. 

Finally, the presented results - obtained with rigorous quantum transport simulations based on an atomistic tight binding Hamiltonian - can be used as input for semiclassical modeling of electron devices, where interband tunneling has to be represented with a synthetic scattering rate.

Support from the EC through the FP7 STEEPER Project (contract n. 257267) is gratefully acknowledged.


\end{document}